\documentclass[a4paper,10pt]{article}

\usepackage{contribution}

\newcommand{\weblink}[2][]{%
    \ifthenelse{\equal{#1}{}}%
    {\textnormal{\url{#2}}}%
    {\textnormal{\href{#2}{#1}}}%
}

\def\beq{\begin{equation}}
\def\eeq#1{\label{#1}\end{equation}}
\def\eeqn{\end{equation}}

\def\beqa{\begin{eqnarray}}
\def\eeqa#1{\label{#1}\end{eqnarray}}
\def\eeqan{\end{eqnarray}}

\let\bar=\overbar

\def\Dslash{\not{\hbox{\kern-4pt $D$}}}
\def\dslash{\not{\hbox{\kern-2pt $\del$}}}

\def\ee{e^+e^-}

\def\msb{{\bar{\ssstyle M \kern -1pt S}}}

\newcommand{\contribution}[7][]{%
  \clearpage
  \thispagestyle{plain}

  \ifthenelse{\equal{#1}{}}
  {\hypersetup{pdftitle={#2}}}
  {\hypersetup{pdftitle={#1}}}
  \hypersetup{pdfauthor={{#3} {#4}}}
  {\centering\normalfont\LARGE\bfseries\sffamily #2 \par\nobreak}
  \lhead{}
  \chead{%
    \textit{\footnotesize XXIInd International Workshop ``High-Energy Physics and Quantum Field Theory'', 
June 24 -- July 1, 2015, Samara, Russia}%
  }
  \rhead{}
  \bigskip
  \begin{center}
    {#3} {#4}\ifthenelse{\equal{#6}{}}{}{\footnote{\weblink[#6]{mailto:#6}}}
    \ifthenelse{\equal{#7}{}}{}{#7} \\
    \textit{#5}
  \end{center}
  \bigskip
}

\renewcommand{\abstract}[1]{%
  \begin{center}
    \begin{minipage}{0.85\textwidth}
      \begin{footnotesize}
        #1
      \end{footnotesize}
    \end{minipage}
  \end{center}
  \bigskip
}

\newcommand{\PCPV}{
\begin{picture}(22,10)
\put(8,-2){\line(2,1){12}}
\put(0,0){$P_{CP}$}
\end{picture}}
\newcommand{\la}{\lambda}
\def\be{\begin{equation}}
\def\ee{\end{equation}}
\def\dd{\displaystyle}
\def\bea{\begin{eqnarray}}
\def\eea{\end{eqnarray}}
\def\nn{\nonumber}
\def\bc{\begin{center}}
\def\ec{\end{center}}
\def\beq{\begin{equation}}
\def\eeq{\end{equation}}
\begin{document} 

\contribution[Recent developments in Neutrino Physics]  
{Recent developments in Neutrino Physics}  
{Davide}{Meloni}  
{Dipartimento di Matematica e Fisica, \\
Universit\'a di Roma Tre\\
Via della Vasca Navale 84, 00144 Rome, Italy}  
{meloni@fis.uniroma3.it}  
 {}  
%

\abstract{%
We discuss the recent progress in neutrino physics in the following fields:
(i) interpretation of the short and long neutrino oscillation data in terms of neutrino flavor transitions; 
(ii) models for neutrino masses and mixings.
}
%

\section{Introduction}
 The main recent interesting result on neutrino mixing has been the measurement of $\theta_{13}$ 
  by T2K\cite{Abe:2011sj}, MINOS\cite{Adamson:2011qu}, DOUBLE CHOOZ\cite{Abe:2011fz}, RENO \cite{Ahn:2012nd} and 
  DAYA-BAY \cite{An:2012eh} experiments.
  Global fits of the oscillation parameters \cite{Fogli:2012ua}, \cite{GonzalezGarcia:2012}, \cite{Tortola:2012te},  
  summarized in Tab.\ref{tabella} for the normal ordering of the neutrino mass eigenstates only, show that the combined value of 
  $\sin^2\theta_{13}$ is about 10 $\sigma$ away 
  from zero and  that its central value is rather large, 
  close to the previous upper bound. 
  For the other mixing parameters, there are solid indications of the deviation of $\theta_{23}$ from the maximal value, 
  probably in the first octant and, thanks to the combined T2K and DAYA-BAY data, a tenuous hints 
  for non-zero $\delta$ is starting to appear from the data.  
  \begin{table}[h!]
  \begin{center}
  \begin{tabular}{|c|c|c|c|}
    \hline
    Quantity & Ref. \cite{Fogli:2012ua} & Ref. \cite{GonzalezGarcia:2012} & Ref. \cite{Tortola:2012te}\\
    \hline
    $\Delta m^2_{sun}~(10^{-5}~{\rm eV}^2)$ &$7.54^{+0.26}_{-0.22}$ & $7.50^{+0.19}_{-0.17}$  & $7.60^{+0.19}_{-0.18}$\\
    $\Delta m^2_{atm}~(10^{-3}~{\rm eV}^2)$ &$2.43\pm 0.06$ & $2.447\pm 0.0047$  & $2.48^{+0.05}_{-0.07}$\\
    $\sin^2\theta_{12}$ &$0.308\pm 0.017$ & $0.304^{+0.013}_{-0.012}$ & $0.323\pm 0.016$\\
    $\sin^2\theta_{23}$ &$0.437^{+0.033}_{-0.023}$ &  $0.452^{+0.052}_{-0.028}$& $0.567^{+0.032}_{-0.124}$\\
    $\sin^2\theta_{13}~(10^{-2})$ &$2.34^{+0.20}_{-0.19}$ &$2.18\pm0.10$ &$2.26\pm0.12$ \\
    $\delta/\pi$ & $1.39^{+0.38}_{-0.27}$& $0.85^{+0.11}_{-0.19}$ & $1.41^{+0.55}_{-0.40}$ \\ 
    \hline
    \end{tabular}
  \end{center}
  \caption{\label{tabella}\it Fits to neutrino oscillation data.}
  \end{table}
  
The interpretation of neutrino appearance and disappearance data in terms of 3$\nu$ oscillation is quite robust and will be revised in Sect.{\ref{sub:osc}}. 
There exists, however, a handful of short baseline data that do not fit into this scheme and demand, among several possibilities, the 
presence of one or more sterile states. These anomalies and their connections with extended models of neutrino oscillations  will be discussed in 
Sect.{\ref{anomalies}}. 
Although the leptonic CP violating phase $\delta$, the octant of $\theta_{23}$ and the 
mass hierarchy are not completely determined by the data, the results shown in Tab.\ref{tabella} indicate that all other mixing  parameters are very well 
constrained; thus, the necessity of focusing on plausible explanation of these values in theoretical framework beyond the current formulation of 
particle theory is mandatory. Some attempts in this direction will be reviewed in Sect.{\ref{models}}.

\section{The standard formulation of neutrino oscillation}
\label{sub:osc}
The presence of non-zero masses for the light neutrinos 
call for the introduction of the $n \times n$ leptonic mixing matrix, $U\equiv U_{PMNS}$ \cite{Maki:1962mu}, which connects the 
flavor eigenstates with the mass eigenstates: 

\begin{equation}
	\nu_\alpha = \sum_i U_{\alpha i}\nu_i\,,
\end{equation}

\noindent
where $\alpha$ denotes one of the active neutrino flavors ($e,\ \mu$ or $\tau$) 
while $i$ runs over 
the light mass eigenstate labels. Being a unitary matrix, $U$ contains, after 
suitable rotations of the leptonic fields, 1/2 $n \cdot (n-1)$ physical angles and 
1/2 $(n-1) \cdot (n-2)$ physical phases.
The neutrino mass differences ($\Delta m^2_{ij}$) and the mixing parameters
($\theta_{ij}, \; \delta$) can be probed by 
studying oscillations between different flavors of neutrinos, 
as a function of the neutrino energy $E$ and the traveled distance 
 $L$. In fact, the oscillation probability 
$P(\nu_\alpha \rightarrow \nu_\beta)$ 
is given by the absolute square of the overlap of 
the observed flavor state, $|\nu_\beta\rangle$, with the time-evolved
initially-produced flavor state, $|\nu_\alpha \,(t)\rangle$.  In vacuum, the 
evolution operator involves just the Hamiltonian $H_0$ of a free particle, 
yielding the well-known result:
\begin{equation}
\begin{array}{rl}
	P(\nu_\alpha \rightarrow \nu_\beta) =&\left|\langle\nu_\beta | 
		e^{-iH_0L}|\nu_\alpha\rangle\right|^2 
	      =	\sum_{i,j} U_{\alpha i}U^*_{\beta i}U^*_{\alpha j}U_{\beta j}
		e^{-i\Delta m^2_{ij}L/2E}\\[0.1in]
	=&P_{\rm CP-even}(\nu_\alpha \rightarrow \nu_\beta) 
		+ P_{\rm CP-odd}(\nu_\alpha \rightarrow \nu_\beta) \; ,
		\\[0.1in]
		\label{eq:prob}
\end{array}
\end{equation}\noindent
where we used the standard approximation that $|\nu\rangle$ is a plane wave, 
$|\nu_i(t)\rangle=e^{-i \,E_i t}|\nu_i(0)\rangle$ and the fact that neutrinos 
are relativistic:  
\begin{equation}
{ E_i}=\sqrt{{p_i^2}+{ m_i^2}}\simeq 
{ p_i}+\frac{{m_i^2}}{2{E_i}} \; .
\end{equation}
The CP-even and CP-odd contributions are:
\begin{equation}
\begin{array}{rl}
	P_{\rm CP-even}(\nu_\alpha \rightarrow \nu_\beta) =&P_{\rm CP-even}(
		\bar{\nu}_\alpha \rightarrow \bar{\nu}_\beta)\\[0.1in]
  	=&\delta_{\alpha\beta} -4\sum_{i>j}\ Re\ (U_{\alpha i}
		U^*_{\beta i}U^*_{\alpha j}U_{\beta j})\sin^2 
		\left({{\Delta m^2_{ij}L}\over{4E}}\right)\\[0.1in]
	P_{\rm CP-odd}(\nu_\alpha \rightarrow \nu_\beta) =&-P_{\rm CP-odd}(
		\bar{\nu}_\alpha \rightarrow \bar{\nu}_\beta)\\[0.1in]
        =&2\sum_{i>j}\ Im\ (U_{\alpha i}U^*_{\beta i}U^*_{\alpha j}
          U_{\beta j})\sin \left({{\Delta m^2_{ij}L}\over{2E}}\right)\\[0.1in]
\end{array}
\label{cprels}
\end{equation}
so that
\begin{equation}
P(\bar\nu_\alpha \to \bar\nu_\beta)= P(\nu_\beta \to \nu_\alpha) = 
P_{\rm CP-even}(\nu_\alpha \rightarrow \nu_\beta) -
P_{\rm CP-odd}(\nu_\alpha \rightarrow \nu_\beta)
\label{cprels2}
\end{equation}
where, by CPT invariance, $P(\nu_\alpha \to \nu_\beta) = 
P(\bar\nu_\beta \to \bar\nu_\alpha)$. 
In eq.(\ref{cprels}),
 $\Delta m^2_{ij} = m(\nu_j)^2 - m(\nu_i)^2$, and the combination 
$\Delta m^2_{ij}L/(4E)$ in $\hbar = c = 1$ units can be replaced
by $1.27 \, \Delta m^2_{ij}L/E$ with $\Delta m^2_{ij}$ 
in ${\rm eV^2}$ and $(L,\ E)$ in $({\rm Km,\ GeV})$. 

For $\beta = \alpha$ (disappearance experiments) no CP-violation 
can appear since the product of the mixing matrix
elements is real. On the contrary, for $\beta \ne \alpha$ (appearance experiments), a CP-violating term could be observed. However,
at distances  $L$ large compared to all the individual oscillation lengths, 
$\lambda_{ij}^{\rm osc} \sim E/\Delta m^2_{ij}$, the sine
squared terms in $P_{\rm CP-even}$ average to 0.5 whereas the sine terms in
$P_{\rm CP-odd}$ average to zero. 
Therefore CP violating effects are largest and hence easiest to observe only
at distances in between the smallest (where the sinus term in $P_{\rm CP-odd}$ can 
be expanded in a series power of its argument) and the largest oscillation lengths.
 
 Neutrino oscillations in matter may differ from oscillations in vacuum
significantly. The effects of the neutrino interactions with matter 
manifest themselves in the oscillation probability in the Mikheyev - Smirnov - 
Wolfenstein (MSW) effect 
\cite{Wolfenstein:1977ue}: even if the mixing angle $\theta$ in vacuum is
very small, matter can enhance neutrino mixing and the probabilities of 
neutrino oscillations if parameters are carefully chosen. This effect can be 
very important for neutrinos
propagating in the Sun and in the Earth as well as for neutrino oscillations in 
supernovae.
Matter effects could be due to the fact that neutrinos can be absorbed 
by the matter constituents, or scattered off them, changing their momentum
and energy.  Neutrinos can also experience forward 
scattering, an elastic scattering in which their momentum is not changed. 
This process is coherent and creates mean potentials $V_a$ for neutrinos 
which are proportional to the number of the scatterers.
 
 The effective Hamiltonian that describes the interactions of neutrinos in
 matter takes into account
 the neutral current (NC) and the charged current (CC) interactions with protons
 and neutrons, resulting in the following expressions for the effective potentials \cite{GonzalezGarcia:2002dz}:
\begin{equation}
V_e = \sqrt{2}\, G_F\,\left(N_e-\frac{N_n}{2}\right)\,,\quad\quad 
V_\mu = V_\tau=\sqrt{2}\, G_F\,\left(-\frac{N_n}{2}\right)\,
\label{Va}
\end{equation}
where $N_e$ and $N_n$ are respectively the electron and neutron number density.

The common
term due to NC interactions appears as a diagonal entry in the interaction 
Hamiltonian (as it is clear from eq.(\ref{Va})), and therefore it does not 
contribute to the flavor states evolution.
The evolution equation for the system can be written as follows\footnote{The 
constant term $p+\frac{m^2_1}{2\,E_\nu}$ appears in every diagonal entries of 
the evolution operator and it does not affect the oscillation probabilities.}:
\begin{equation}
i{d\nu_\alpha\over dt} = \sum_\beta \left[\left(p+\frac{m^2_1}{2\,E_\nu}\right)\, \delta_{\alpha
\beta}+\left( \sum_{j>1} U_{\alpha j} U_{\beta
j}^* {\Delta m_{1j}^2\over 2E_\nu} \right) + A\, \delta_{\alpha e}
\delta_{\beta e} \right] \nu_\beta \,,  \label{eq:prop}
\end{equation}
where we introduced the constant
\begin{equation}
A = \sqrt2 G_F N_e  = 0.76 \times 10^{-4}{\rm\,eV^2} \, Y_e\,
\rho({\rm\,g/cm^3})\,
\label{eq:defnA}
\end{equation}
(for $\bar{\nu_e}$ A is replaced with -A). 
$Y_e$ is the electron fraction and $\rho$ is the matter density. 
Density profiles through the 
Earth can be calculated using the Preliminary Earth Model (PREM) \cite{prem}. 
For
neutrino trajectories through the Earth's crust, the density is typically of
order 3~gm/cm$^3$ ($Y_e \simeq 0.5$) and it can be considered as a constant.
For very long baselines, however, this approximation is not sufficient
and we must explicitly take account of $\rho$.
%

For the case of three neutrino families, 
 the lepton mixing matrix $U$ contains three mixing angles 
$\theta_{12}$, $\theta_{13}$ and $\theta_{23}$ and one CP-violating phase 
$\delta$ \footnote{In the case of Majorana 
neutrinos there are two additional phases but they do not affect the
flavor transitions.} and can be built from three rotation matrices on
different subspaces:
\begin{equation}
U=U_{23}(\theta_{23})\,U_{13}(\theta_{13}, \delta)\,U_{12}(\theta_{12})
\end{equation}
leading to its standard parameterization \footnote{Notice that the 
parameterizations differing in the position of the $\delta$ phase are 
physically equivalent.}:
\begin{equation}
U =
\left(\begin{array}{ccc}
c_{12}\,c_{13}   & s_{12}\, c_{13}    & s_{13}\, e^{-i\delta} \\
-s_{12}\,c_{23}-c_{12}\,s_{23}\,s_{13}\,e^{i\delta} &
c_{12}\,c_{23}-s_{12}\,s_{23}\,s_{13}\,e^{i\delta} &
s_{23}\,c_{13}   \\
s_{12}\,s_{23}-c_{12}\,c_{23}\,s_{13}\,e^{i\delta} &
-c_{12}\,s_{23}-s_{12}\,c_{23}\,s_{13}\,e^{i\delta} &
c_{23}\,c_{13}
\end{array}
\right)\,, 
\label{U3}
\end{equation}
where $c_{ij}=\cos \theta_{ij}$, $s_{ij}=
\sin\theta_{ij}$. 
Assuming that the
$\nu_3$ is the neutrino eigenstate that is separated from the other two,
the sign of $\Delta m^2_{13}$ can be either positive or
negative, corresponding to the case where $\nu_3$ is either above or
below, respectively, the other two mass eigenstates. The magnitude of
$\Delta m^2_{13}$ determines the oscillation length of atmospheric
neutrinos, while the magnitude of $\Delta m^2_{12}$ determines the
oscillation length of solar neutrinos, and thus $|\Delta m^2_{12}| \ll
|\Delta m^2_{13}|$ (see Tab.\ref{tabella}).

 The oscillation probabilities between various 
flavor states can be obtained from the general expression 
in eq.(\ref{cprels}) and are generally quite cumbersome.
In the approximation that we neglect oscillations 
driven by the small $\Delta m_{12}^2$ scale, only the CP-even part of the transition
probabilities survives and their expression is relatively simple. As examples we show
$P(\nu_e \rightarrow \nu_e)$, which is especially useful for extracting the value of the reactor angle and 
$P(\nu_\mu \rightarrow \nu_\tau)$, from which a good knowledge of the atmospheric mass and mixing angle can be obtained: 
\begin{eqnarray}
P(\nu_e \rightarrow \nu_e) \simeq & 1 
-4|U_{e3}|^2 (1 - |U_{e3}|^2)\nonumber
\sin^2 \Bigl (\frac{\Delta m^2_{atm}L}{4E} \Bigr ) \\[0.05in]
= & 1 - \sin^2(2\theta_{13}) 
\sin^2 \Bigl (\frac{\Delta m^2_{atm}L}{4E} \Bigr ) \; ,\nonumber
\end{eqnarray}

\begin{eqnarray} 
P(\nu_\mu \rightarrow \nu_\tau) \simeq & \nonumber
4|U_{\mu 3}|^2 |U_{\tau 3}|^2 
\sin^2 \left({{\Delta m^2_{atm} L} \over{4E}}\right) \\[0.05in]
 =& \sin^2(2\theta_{23})\cos^4(\theta_{13}) 
\sin^2 \Bigl (\frac{\Delta m^2_{atm}L}{4E} \Bigr ) \;. \label{pnuenutau}
\end{eqnarray}

The CP-even transition probabilities for antineutrinos can be obtained from the first 
relation in eq.(\ref{cprels}). 

The evaluation of the CP-odd part of the transition
probabilities is a little bit more complicated due to the combined presence of
two mass differences. However it is possible to obtain very simple expressions
expanding the transition probabilities to second order in the small parameters 
$\theta_{13}$, $\Delta m_{12}^2 / \Delta m_{23}^2$ and 
$\Delta m_{12}^2 L /2 E$ \cite{Cervera:2000kp}, \cite{Donini:2002rm}:
\begin{eqnarray} 
\PCPV(\nu_e \rightarrow \nu_\mu)  = - \PCPV(\nu_e \rightarrow \nu_\tau) 
= - \PCPV(\nu_\mu \rightarrow \nu_\tau) = \nonumber
\end{eqnarray}
\begin{eqnarray} 
\tilde{J}\;\frac{\Delta m_{12}^2 L}{4 E} \, \sin \delta \,
\sin^2\left(\frac{\Delta m_{13}^2 L}{4 E}\right) 
\end{eqnarray}
where $\tilde{J} = c_{13} \, \sin 2 \theta_{12}\, \sin 2 \theta_{23} \, 
\sin 2 \theta_{13}$ (see eq.(\ref{cprels}) for antineutrinos).

The interaction of neutrinos with matter is easily described by means of 
eq.(\ref{eq:prop}); in the approximation where we neglect oscillations 
driven by the small $\Delta m_{12}^2$ scale, the evolution equations are:

\begin{equation}
i {d\over dt}
\left( \begin{array}{c} \nu_e \\ \nu_\mu \\ \nu_\tau  \end{array} \right)
= {\Delta m^2\over 2E_\nu}
\left( \begin{array}{ccc}
{2 E_\nu \over \Delta m^2}\,A + |U_{e3}|^2 & U_{e3}U_{\mu3}^* & U_{e3}U_{\tau3}^* \\
U_{e3}^*U_{\mu3} & |U_{\mu3}|^2 & U_{\mu3}U_{\tau3}^* \\
U_{e3}^*U_{\tau3} & U_{\mu3}^*U_{\tau3} & |U_{\tau3}|^2
\end{array} \right)
\left( \begin{array}{c} \nu_e \\ \nu_\mu \\ \nu_\tau \end{array} \right)
\,.
\end{equation}

The structure of the mixing matrix assumes a simple form for matter of 
constant density; in that case, the flavor eigenstates are
related to the mass eigenstates $\nu_j^m$ by
\begin{equation}
\nu_\alpha = \sum U_{\alpha j}^m | \nu_j^m \rangle \,,
\end{equation}
and
\begin{equation}
U^m = \left( \begin{array}{ccc}
      0 &  c_{13}^m & s_{13}^m \\
-c_{23} & -s_{13}^m s_{23} & c_{13}^m s_{23} \\
 s_{23} & -s_{13}^m c_{23} & c_{13}^m c_{23}
\end{array} \right)
\label{eq:matter}
\end{equation}
The resonance condition is obtained by looking at the relation 
\footnote{The case $\theta_{13}^m = 
\theta_{13}$ is recovered in the limit A $\to$ 0.}:
\begin{equation}
\sin^2 2\theta_{13}^m = \sin^22\theta_{13} /
\left[{ \left( {2 E_\nu \, A\over\Delta m^2} - \cos 2\theta_{13} \right)^2
+ \sin^2 2\theta_{13}} \right] . 
\label{eq:sin}
\end{equation}
Then there is an enhancement when
\begin{equation}
A = \frac{\Delta m^2}{2 E_\nu} \cos2\theta_{13}
\end{equation}
or equivalently
\begin{equation}
E_\nu \approx 15{\rm\ GeV} \left(\Delta m^2 \over 3\times
10^{-3}{\rm\,eV^2}\right) \left( 1.5{\rm\ g/cm^3}\over \rho Y_e \right)
\cos2\theta_{13} \,. \label{eq:Enu}
\end{equation}

The resonance occurs only for positive $\Delta m^2$ for neutrinos 
and only for negative $\Delta m^2$ for anti-neutrinos.
For negative
$\Delta m^2$ the oscillation amplitude in eq.(\ref{eq:sin}) is smaller than the
vacuum oscillation amplitude. Thus the matter effects give us a way in
principle to determine the sign of $\Delta m^2$.
It is important to stress that matter effects for long
baseline experiments become
important for path-lengths greater 
than 2000 Km.    

\section{Anomalies in neutrino oscillation}
\label{anomalies} 
\begin{figure}[h!]
\centering
\includegraphics[width=5.5cm]{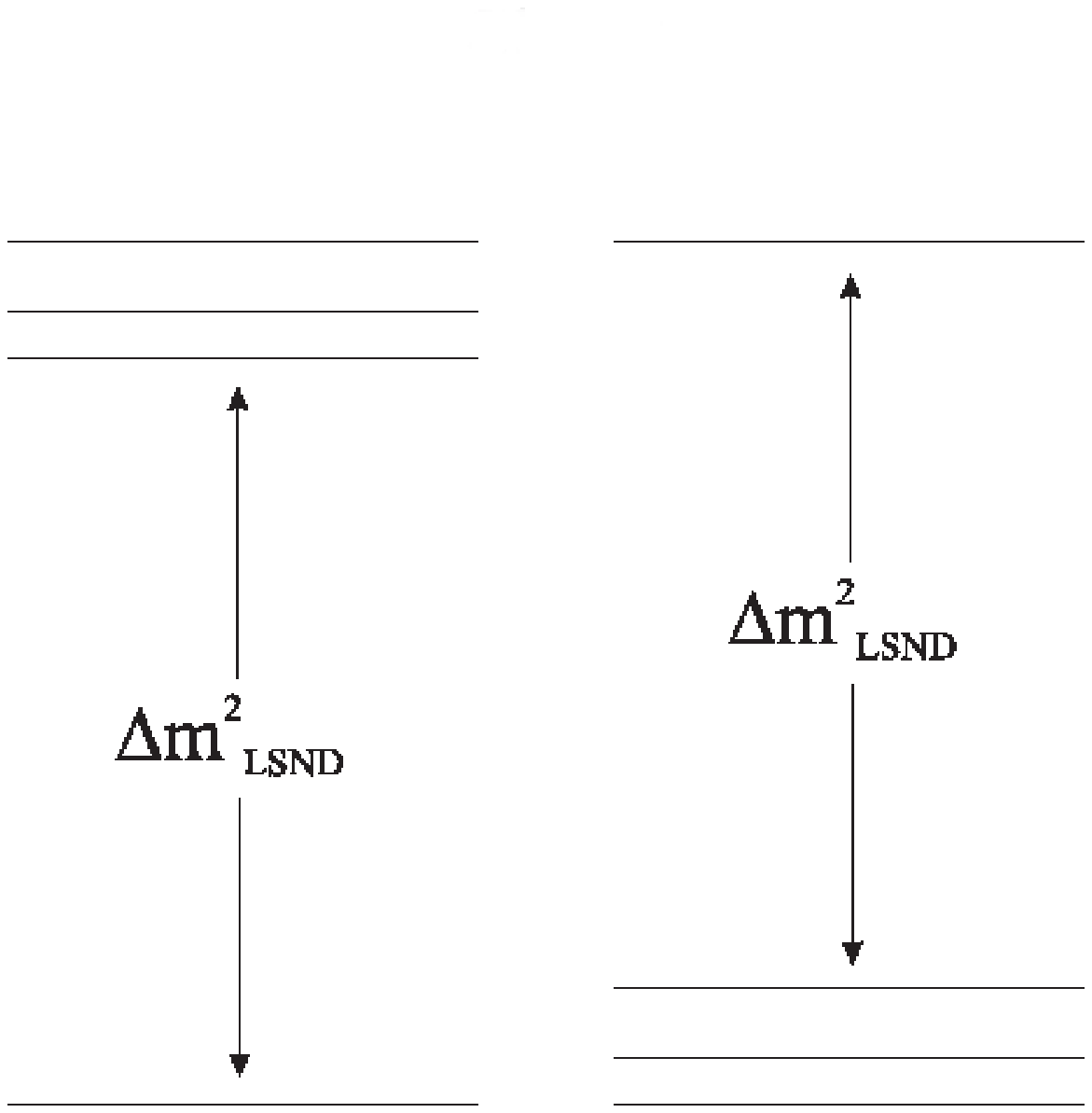} 
\vspace*{0.5cm}
\caption{{\it Different types  mass spectrum in the  3+1
scenario.}}
\label{fig:classes}
\end{figure}

In the last years a certain number of hints have been collected in neutrino oscillation experiments which pointed towards the 
existence of sterile neutrinos \cite{white}, 
that is neutrinos with no weak interactions.
The LSND results \cite{Athanassopoulos:1995iw} published data showing candidate events that are 
consistent with $\bar\nu_\mu\to \bar\nu_e$ oscillations at very short distance.
Data could be explained as if three almost degenerate (mainly active) 
neutrinos, accounting for the solar and atmospheric oscillations, 
were separated from the fourth (mainly sterile) one by the large
LSND mass difference, $\Delta m^2_{LSND}$, a situation called {\it 3+1 scheme}, see Fig.(\ref{fig:classes}).
The other possibility, that is two almost degenerate neutrino pairs, accounting respectively for the
solar and atmospheric oscillations, separated by the LSND mass gap \cite{Donini:2001xy}, is already ruled out \cite{Maltoni:2002xd}.

For $n = 4$, the PMNS matrix contains six independent rotation angles $\theta_{ij}$ 
and three (if neutrinos are Dirac fermions) or six (if neutrinos are Majorana
fermions) phases $\delta_i$. 
This large parameter space is actually reduced to a smaller subspace whenever some of the 
mass differences become negligible. Consider for example the measured hierarchy in the 
mass differences, 
\begin{equation}
\Delta m^2_{sol} \ll \Delta m^2_{atm} \ll \Delta m^2_{LSND} \, , 
\end{equation}
and define
\begin{equation}
\Delta_{ij} = \frac{\Delta m^2_{ij} L}{4 E_\nu} \, ;
\label{def:deltaij}
\end{equation}
then, at short distance of $L = O(1)$ Km, for neutrinos up to $O (10)$ GeV, the ``one-mass dominance'' sets in,
\begin{eqnarray}
\Delta_{sol} \; , \; \Delta_{atm} \ll 1 \, , \nonumber \\
\Delta_{LSND} = O(1) \, 
\end{eqnarray}
and the whole three-dimensional subspace $(1-2-3)$
is irrelevant for short-baseline oscillation experiments; the physical parameter
space now contains just three rotation angles and no phases. This means that the (otherwise) complicated
neutrino transitions can be approximated by formulae depending on only one relevant mass difference and mixing angle, 
in the form:
\begin{eqnarray}
\label{prob3p1}
 P(\nu_\alpha \to \nu_\beta) = \delta_{\alpha\beta}-(2 \delta_{\alpha\beta}-1 ) \sin^2 2\theta_{\alpha\beta} 
 \sin^2 \left(\frac{\Delta m^2_{ij}L}{4E_\nu}\right) \,,
\end{eqnarray}
with the meaning that
\begin{eqnarray}
 \sin^2 2\theta_{e\mu} &=& 4 |U_{e4}|^2  |U_{\mu 4}|^2 {\rm \; \;\;\; \; \;\;\;\;\, \;for}\; \nu_\mu \to \nu_e\; {\rm appearance} 
 \nonumber \\
 \sin^2 2\theta_{e e} &=& 4 |U_{e4}|^2  (1-|U_{e 4}|^2 {\rm \;\;\;\, for}\;  \nu_e\; {\rm disappearance} \label{angles}\\
 \sin^2 2\theta_{\mu\mu} &=& 4 |U_{\mu 4}|^2  (1-|U_{\mu 4}|^2) {\rm \; for}\; \nu_\mu\;  {\rm disappearance.} \nonumber
\end{eqnarray}

Notice that when considering CP-violating phenomena at least two mass
differences should be taken into account (one neglects the solar mass difference and considers the atmospheric
one as a perturbation): this is called ``two-mass dominance''
approximation. Regardless of the scheme, 
the parameter space contains 5 angles and 2 phases. 

Anomalies in neutrino oscillation data belong to the following cathegories:
\begin{itemize}
\item the LSND results \cite{Athanassopoulos:1995iw} provide the only positive signature of oscillations in accelerator experiments. 
The data showed candidate events 
consistent with $\bar\nu_\mu\to \bar\nu_e$ oscillations and obtained further supporting 
evidence by the signal in the $\nu_\mu\to\nu_e$ channel. 
Updated results including the runs till 1998
fixed the total fitted excess at 
$87.9\pm22.4\pm6$ 
events, corresponding to an oscillation probability of 
$(2.64\pm0.67\pm0.45)\times10^{-3}$ or, in terms of oscillation parameters, to $\Delta m^2_{LSND}=1.2$ eV$^2$ and $\sin^22\theta=0.003$;
\item a combined analysis of  $\nu_\mu \to \nu_e$  together with $\bar{\nu}_\mu \to \bar{\nu}_e$ data has been published by the 
MiniBooNE experiment \cite{Mboo:2012}; 
they observed an excess of events over the expected backgrounds in the low energy  region part of the spectrum, below ~500 MeV. 
The allowed region from MiniBooNE anti-neutrino data partially overlaps with the region preferred by the LSND data \cite{LSND};
\item anomalies in Gallium experiments (SAGE \cite{Abdurashitov:1998ne} and GALLEX \cite{Hampel:1997fc}): 
they measured an electron neutrino flux from the Sun smaller than expected \cite{Acero:2007su, Giunti:2010zu}
($\nu_e$ disappearance experiments);
\item anomalies due to new computations of reactor neutrino fluxes \cite{Mention:2011,Huber:2011}: fluxes from reactor neutrinos are 
$\sim 3.5\%$ larger than in the past
\cite{Vogel:1980bk,VonFeilitzsch:1982jw,Schreckenbach:1985ep, Hahn:1989zr}, so that 
experiments with L$\le$ 100 m show deficit of neutrinos (short-baseline $\nu_e$ disappearance experiments, like Bugey \cite{Declais:1994ma}, Rovno \cite{Kuvshinnikov:1990ry},
Palo Verde \cite{Boehm:2001ik}, DoubleChooz \cite{Abe:2012tg}...).
\end{itemize}

In addition there are {\it null results} experiments, which gave no signal. Among them:
$\nu_\mu$ disappearance (CDHS \cite{Dydak:1983zq}, Super Kamiokande \cite{Bilenky:1999ny}, MINOS \cite{Adamson:2010wi}) and $\nu_e$
appearance (KARMEN \cite{Armbruster:2002mp}, NOMAD \cite{Astier:2003gs}, ICARUS \cite{Antonello:2012}, OPERA \cite{agafo}).

The previous data can be analyzed in the 3+1 scheme and adopting the two-flavor probabilities given in eq.(\ref{prob3p1}). 
Global fits of {\it $\nu_e$ appearance} data are consistent among themselves and 
give a best fit point in $(\sin^2 2\theta_{e\mu},\Delta m^2_{41})=(0.013,0.42~ eV^2)$ \cite{kopp}; also the  
{\it $\nu_e$ disappearance} data are not in contradiction and point toward a best fit in 
$(\sin^2 2\theta_{e e},\Delta m^2_{41})=(0.09,1.78 ~eV^2)$. 
On the other hand, experiments
probing $\bar \nu_\mu$ disappearance have not reported any hints for a positive signal and this gives strong constraints 
on the matrix element $|U_{\mu 4}|$, that is on $\sin^2 2\theta_{\mu\mu}$. Since the three mixing angles entering eq.(\ref{angles})
are related by
\begin{eqnarray}
 \sin^2 2\theta_{e\mu} \sim \frac{1}{4} \sin^2 2\theta_{e e} \sin^2 2\theta_{\mu\mu}\,,
\end{eqnarray}
one would have expected a quadratic suppression on the appearance amplitude, in contrast with the obtained best fit 
value  $\sin^2 2\theta_{e\mu}=0.013$. Roughly speaking, this is the source of the well-known tension between appearance and 
disappearance experiments. To quantify the consistency of different parts of
the global data, one can use the so-called parameter goodness of fit test; 
the authors of Ref.\cite{kopp} found a value around $10^{-4}$, which strongly indicates
the poor agreement of the two sets of data.

Notice that the 3+1 fit is much improved if 
the low energy MiniBooNE data are not included \cite{laveder}. In fact,  this subset of the data is 
incompatible with neutrino oscillations since they require a small value of $\Delta m^2_{14}$ and a large value 
of $\sin^2 2\theta_{e\mu}$ well beyond what is allowed by the data of other experiments. Removing the 
the low energy excess of MiniBooNE from the fit, the authors of Ref.\cite{laveder} found a parameter goodness of fit as large as $9\%$.

In conclusion, the situation is at present quite confuse and additional experimental effort is needed to establish the 
existence of sterile neutrinos.

 \section{Models of masses and mixing}
 \label{models}
 \subsection{Models based on discrete symmetries}
 Looking at the results of Tab.\ref{tabella}, one is still tempted to recognize some special mixing patterns 
 as good first approximations to describe the data, the most famous ones being the Tri-Bimaximal (TB \cite{Harrison}), 
 the Golden Ratio (GR \cite{GR1}) and the 
 Bi-Maximal (BM) mixing. The corresponding mixing matrices all have:
 \begin{equation}
 \sin^2{\theta_{23}}=1/2,\qquad \sin^2{\theta_{13}}=0 
 \end{equation}
 and differ by the value of the solar 
 angle $\sin^2{\theta_{12}}$, which is:
 \begin{equation}
 \sin^2{\theta_{12}}=1/3 {\rm\; for\; TB}, \qquad \sin^2{\theta_{12}} = \frac{2}{5+\sqrt{5}}\sim 0.276 {\rm\; for\; GR}, \qquad 
  \sin^2{\theta_{12}} =\frac{1}{2}{\rm\; for\; BM}\,.
 \end{equation}
 
 Being a leading order approximations (LO), all previous patterns  need corrections 
 (for example, from the  diagonalization of charged leptons) to describe the current mixing angles. In particular, the relatively 
 large value of the reactor angle requires sizable corrections of the order of the Cabibbo angle $\lambda_C$, for all three patterns; 
 on the other hand, the deviations 
 from the LO values of $\sin^2{\theta_{12}}$ must be small enough in 
 the TB and GR cases but as large as $\lambda_C$ for the BM pattern. Finally, corrections not too much larger than 
 $\lambda_C^2$ can be tolerated by $\sin^2\theta_{23}$.
 Since the corresponding mixing matrices have the form of rotations with special angles,  
 discrete flavor groups naturally emerge as good candidates. The most studied groups have been the 
 permutation groups of four object, $S_4$ and $A_4$, see Ref.~\cite{Altarelli:2010gt} for an exhaustive review and 
 \cite{Altarelli:2006ty} for examples involving $A_4$.
 The important point for model building is that these symmetries must be broken by suitable scalar fields $\varphi$ that 
 take a vacuum expectation value (vev) at large scale $\Lambda$, so they generally provide a new adimensional parameter 
 $\xi = \langle \varphi\rangle/\Lambda$. The breaking must preserve different subgroups in the charged lepton and 
 neutrino sectors, otherwise the neutrino mixing matrix would be the identity matrix and no mixing will be generated.
 The desired directions in flavor space are generally difficult to achieve, so consistent models are those where 
 the vevs of the scalar fields can be naturally obtained from the minimization of the scalar potential.

 Beside the models based on TB, one can consider models where  BM mixing holds in the neutrino sector at LO and the 
 relatively large corrective terms for $\theta_{12}$ and $\theta_{13}$, of $\mathcal{O}(\lambda_C)$, arise from the diagonalization of 
 charged lepton masses; the atmospheric angle, however, should deviate from maximal mixing by quantities not much larger than 
 $\mathcal{O}(\lambda_C^2)$.
 Explicit models of this type based on the group $S_4$ have been developed in Ref.~\cite{Altarelli:2009gn}.

For the BM mixing,
the mixing matrix has the form
\be
U_{BM}= \left(
\begin{array}{ccc}
\dd\frac{1}{\sqrt 2}&\dd-\frac{1}{\sqrt 2}&0\\
\dd\frac{1}{2}&\dd\frac{1}{2}&-\dd\frac{1}{\sqrt 2}\\
\dd\frac{1}{2}&\dd\frac{1}{2}&\dd\frac{1}{\sqrt 2}
\end{array}
\right)\;,
\label{BM}
\ee
corresponding to the following mass matrix:
\be
m_{\nu BM}=\left(
\begin{array}{ccc}
x&y&y\\
y&z&x-z\\
y&x-z&z
\end{array}
\right)\;,
\label{gl2}
\ee
where $x$,$y$ and $z$ are three complex numbers.
One can consider the possibility that BM is the mixing in the neutrino sector and that the  rather large
corrective terms to $\theta_{12}$ and $\theta_{13}$ arise from the diagonalization of the charged lepton mass matrix, as 
 obtained in models 
based on the discrete symmetry $S_4$ \cite{Altarelli:2009gn,Meloni:2011fx}.
This idea is in agreement with the well-known empirical 
quark-lepton complementarity relation \cite{Raidal:2004iw}-\cite{Antusch:2005ca}, 
$\theta_{12}+\theta_C\sim \pi/4$, 
where $\theta_C$ is the Cabibbo angle or, to be less optimistic, with the ``weak'' complementarity relation
$\theta_{12}+\mathcal{O}(\theta_C)\sim \pi/4$. In addition, the measured value of 
$\theta_{13}$ is itself of order $\theta_C$:  $\theta_{13}\sim \theta_C/\sqrt{2}$.  

In the following, two examples of GUT models based on BM will be considered \cite{Altarelli:2015foa}: 
one is based on $SU(5)$ \cite{Meloni:2011fx} and realizes the program of imposing
the BM structure in the neutrino sector and then correcting it by terms arising from the diagonalization of charged lepton masses.
The other is an $SO(10)$ model based on Type-II see-saw \cite{blankenburg}, where the origin of BM before diagonalization of charged leptons 
is left unspecified. 

In the first case we deal with a variant of the SUSY $SU(5)$ model in 4+1 
dimensions with a flavor symmetry $S_4 \otimes Z_3 \otimes U(1)_R \otimes U(1)_{FN}$ \cite{Altarelli:2009gn,Meloni:2011fx}, 
where $U(1)_R$ implements the 
R-symmetry while $U(1)_{FN}$ is a Froggatt-Nielsen (FN) symmetry \cite{Froggatt:1978nt} that induces the hierarchies of fermion masses 
and mixings. 
The particle assignments are displayed in Tab.\ref{tab:Multiplet1}. 
\begin{table}[h!]
\begin{center}
\begin{tabular}{|c||c|c|c|c|c|c|c|c|c|c|c|c|c|c|c|c||}
\hline
{\tt Field} & $F$ & $T_1$ & $T_2$ & $T_3$ & $H_5$ & $H_{\overline 5}$ &   $\varphi_\nu$ & $\xi_\nu$ &  $\varphi_\ell$& $\chi_\ell$ &  
$\theta$ &  $\theta^\prime$ & 
$\varphi^0_\nu$& $\xi^0_\nu$ &  $\psi^0_\ell$& $\chi^0_\ell$ 
\\
\hline\hline
SU(5) & $\bar{5}$  & 10 & 10 & 10 & 5 & ${\overline {5}}$ &  1 & 1 & 1 & 1 & 1 & 1 & 1 & 1 & 1 & 1  \\
\hline
$S_4$  &  $3_1$ & 1  & 1 & 1 & 1 & 1 &   $3_1$ & 1 & $3_1$ & $3_2$ &  1 & 1 & $3_1$ & 1 & 2 & $3_2$     \\
\hline
$Z_3$ & $\omega$ & $\omega$ &1  &  $\omega^2$ & $\omega^2$ &$\omega^2$ &  1 & 1 & $\omega$ &$\omega$ &1 &  $\omega$ & 1 & 1 & 
$\omega$ & $\omega$  \\
\hline
$U(1)_R$ & 1 & 1 & 1 & 1 & 0 & 0&  0 & 0 & 0 & 0& 0 & 0 & 2 & 2 & 2 & 2 \\
\hline
$U(1)_{FN}$ & 0 & 2 & 1 & 0 & 0 &  0 & 0 & 0 & 0& 0 & -1 & -1 & 0 & 0 & 0 & 0 \\
\hline 
  & {\tt br} &  {\tt bu} & {\tt bu} & {\tt br} &  {\tt bu} &  {\tt bu} &   {\tt br} &  {\tt br} &  {\tt br} & 
 {\tt br} &  {\tt br} &  {\tt br} &  {\tt br} &  {\tt br} &  {\tt br} &  {\tt br}\\
\hline 
\end{tabular}
\caption{\label{tab:Multiplet1}\it Matter assignment of the model. The symbol ${\tt br}({\tt bu})$ indicates that the corresponding 
fields live
on the brane (bulk).}
\end{center}
\end{table}
The first two generation tenplets $T_1$ and $T_2$ and the Higgs $H_5$ and $H_{\bar 5}$
are in the bulk  while all the other ones are 
on the brane at $y=0$; this  introduces some extra hierarchy for some of the couplings \cite{5DSU5}-\cite{Altarelli:2008bg}. 
At leading order (LO) the $S_4$ symmetry is broken 
down to suitable different subgroups in the charged lepton sector and in the neutrino sector by the VEV's of the flavons 
$\varphi_\nu$, $\xi_\nu$,  $\varphi_\ell$ and $\chi_\ell$ (whose proper alignment is implemented  in a natural 
way  by the driving fields $\varphi^0_\nu$,  $\xi^0_\nu$,  $\psi^0_\ell$, $\chi^0_\ell$). The VEVs of the $\theta$ and $\theta^\prime$ 
fields break the FN symmetry.  As a result, at LO the charged lepton masses are diagonal and exact BM is realized for neutrinos.  
Corrections to diagonal charged leptons and to exact BM are induced by vertices of higher dimension in the Lagrangian, suppressed by 
powers of a large scale $\Lambda$. 
We adopt the definitions:
\bea
\frac{v_{\varphi_\ell}}{\Lambda} \sim \frac{v_{\chi}}{\Lambda} 
\sim \frac{v_{\varphi_\nu}}{\Lambda} \sim \frac{v_{\xi}}{\Lambda}\sim
\frac{\langle \theta \rangle}{\Lambda} \sim \frac{\langle \theta^\prime \rangle}{\Lambda} \sim
s\equiv \lambda_C\,,
\label{vsup}
\eea  
where $s=\dd\frac{1}{\sqrt{\pi R \Lambda}}$ is the volume suppression factor. 
It turns out that this simple choice
leads to a good description of masses and mixings.

For the charged lepton masses, the matter assignment of the model gives rise to the following mass matrix:
\be
m_e \sim \left(
\begin{array}{ccc}
a_{11}\lambda^5&a_{21} \lambda^4& a_{31}\lambda^2\\
a_{12} \lambda^4&-c \lambda^3& ...... \\
a_{13}\lambda^4&c \lambda^3&a_{33} \lambda
\end{array}
\right) \lambda ,
\label{melam}
\ee
where all matrix elements are multiplied by generic coefficients  of $\mathcal{O}(1)$.
The corresponding lepton rotation is given by:
\bea
\label{ul}
U_\ell \sim  
\left(
\begin{array}{ccc}
1  &u_{12}\lambda&u_{13}\lambda  \\
-u_{12}^* \lambda&  1 & 0 \\
-u_{13}^* \lambda   &  
-u_{12}^* u_{13}^*\lambda^2     & 1 \\
                     \end{array}
                   \right)\,,
\eea 
($u_{ij}$ again of  $\mathcal{O}(1)$) so that $\theta_{23}^\ell = 0$ in this approximation.

The neutrino sector of the model is unchanged with respect to Ref.\cite{Meloni:2011fx}. At LO, the mass matrix of eq.({\ref{gl2}})
is obtained from the Weinberg operator, so 
the results for the mixing angles are easily derived:
\bea
\sin{\theta_{13}} &=&  \frac{1}{\sqrt{2}} |u_{12}-u_{13}|\lambda   \nn \qquad
\sin^2{\theta_{12}} = \frac{1}{2}- \frac{1}{\sqrt{2}}~Re(u_{12}+u_{13})\lambda   \qquad
\sin^2{\theta_{23}} =\frac{1}{2}+ \mathcal{O}(\lambda^2) \nn\,.
\eea
We see that, with $\lambda \sim \lambda_C$, the model realizes the "weak" complementarity relation and the experimental fact that 
$\sin{\theta_{13}}$ is of the same order 
than the shift of 
$\sin^2{\theta_{12}}$ from the BM value of 1/2, both of order $\lambda_C$. 

The second model presented here is based on the $SO(10)$ gauge group.
In $SO(10)$ the main added difficulty with respect to $SU(5)$ 
is that one generation of  fermion belongs to the 16-dimensional representation, 
so that one cannot take advantage of the properties of the 
$SU(5)$-singlet right-handed neutrinos. A possible strategy to separate charged fermions and 
neutrinos is to assume 
the dominance of type-II see-saw  with respect to the more usual type-I see-saw. 
In models of this type, the neutrino mass formula becomes
\begin{eqnarray}
{\cal M}_\nu~\sim~fv_L\;,
\label{eq4}
\end{eqnarray}
where $v_L$ is the vev of the $B-L=2$ triplet in the ${\bf\overline{126}}$ Higgs
field and $f$ is the Yukawa coupling matrix of the $\bf 16$ with the same ${\bf\overline{126}}$.

For generic eigenvalues $m_i$, the most general matrix that is diagonalized by the BM unitary transformation is given by:
\beq
f=U_{BM}^* {\rm diag}(m_1,m_2,m_3) U_{BM}^\dagger~~~,
\label{numass}
\eeq
where $U_{BM}$ is the BM mixing matrix given in eq.(\ref{BM}). 
However, a similar transformation can also be used with $U_{BM}$ replaced by $U_{TB}$; 
as a result, the matrices $f$ obtained with this two different approaches are related by a change of the charged lepton basis 
induced by a unitary matrix. 
As one could decide to work in a basis where the matrix $f$ is diagonalized by the TB matrix or by BM matrix, 
the result of a fit performed in one basis should lead to the same $\chi^2$ than the fit 
in other basis, so the $\chi^2$ cannot decide whether TB or BM is a better starting point. 
Then we need another "variable" to compare whether the data prefer to start from TB or BM.
One possibility is to measure the amount of fine-tuning needed to fit a set of data; to this aim, a parameter $d_{FT}$ was 
introduced in Ref. \cite{blankenburg}:
\begin{equation}
\label{fine-tuning}
d_{FT} = \sum\mid \frac{par_i}{err_i} \mid ,
\end{equation}
where $err_i$ is the "error" of a given parameter $par_i$ defined as the shift from the best fit value that changes 
the $\chi^2$ by one unit, with all other parameters fixed at their best 
fit values.

\begin{figure}[h!]
\centering
\includegraphics[width=6.5cm]{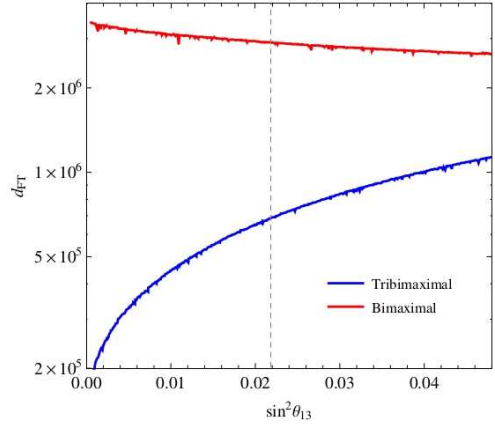} 
\vspace*{0.5cm}
\caption{\label{fig4}\it In the $SO(10)$ model the fine tuning parameter $d_{FT}$ increases (decreases) with $\sin^2{\theta_{13}}$ in the TB 
(BM) cases. For the physical value $\sin^2{\theta_{13}} \sim 0.022$ it is about 4 times larger in the BM case.}
\end{figure}

A study of the fine tuning parameter when the fit is repeated with the same data except for $\sin^2\theta_{13}$, 
which is moved from small to large, shows that the fine tuning increases (decreases) with $\sin\theta_{13}$ for TB (BM), as 
shown in Fig.(\ref{fig4}). 

A closer look at the figure reveals that
both BM and TB scenarios are compatible with the data  for similar values of the fine tuning parameter, 
especially for relatively large $\theta_{13}$. We have also observed that 
high $d_{FT}$ values are predominantly driven by the smallness of the
electron mass  combined with its extraordinary measurement
precision. 

\subsection{Models based on abelian $U(1)$} 
The relatively large value of $\theta_{13}$
 and the fact that $\theta_{23}$ is not maximal both point to the direction of models based on 
 Anarchy \cite{Hall:1999sn,deGouvea:2003xe}, that is the assumption that no special symmetry is needed in the leptonic sector, 
 and that the values of neutrino masses and mixing are reproduced by chance. 
 Anarchy can be formulated in a $U(1)$ context a la Froggatt-Nielsen \cite{Froggatt:1978nt}:  
 a mass term is allowed at the renormalisable level only if the $U(1)_{FN}$  charges add to zero.
 Breaking  the $U(1)_{FN}$ symmetry spontaneously by
 the vevs $v_f$ of a number of flavon fields with non-vanishing charge
 allows to rescue the forbidden vertex, although suppressed by powers of the small parameters $\lambda = v_f/M$, 
 with $M$ a large mass scale. Since these invariant mass terms appear with 
 arbitrary coefficients of order 1, typically the number of parameters exceeds the number of observable quantities and make 
 this kind of models less predictive than the ones based on non-abelian discrete symmetries. 
 
 Opposite to Anarchy, generic $U(1)$ models are characterized by well-defined hierarchies of 
 the neutrino mass matrix elements and are often referred
  to as hierarchical models. 
 The authors of Ref.~\cite{Altarelli:2012ia} have performed an updated analysis of the performance of anarchical versus hierarchical models
 in the $SU(5) \otimes U(1)_{FN}$ context, which also allows to implement a parallel  
 treatment of quarks and leptons. Among the different charge assignments of the 10, $\bar 5$ and the $SU(5)$ singlet, 
 we focus here on two different realizations:
 \begin{itemize}
 \item Anarchy (A): 10=(3,2,0), $\bar 5$=(0,0,0)  1=(0,0,0)
 \begin{equation}
 m_{\ell}= \left(  \begin{matrix}  \la^3 &    \la^3 &   \la^3 \\  \la^2 &    \la^2 &   \la^2 \\ 1 &   1 &   1 \end{matrix}
 \right)\,,\qquad\qquad
  m_{\nu}= \left(
  \begin{matrix}  1 &    1 &   1 \\  1 &   1 &  1 \\ 1 &   1 &   1
  \end{matrix}     
  \right)\,.
  \label{A}
 \end{equation}
 \item Hierarchy (H): 10=(5,3,0), $\bar 5$=(2,1,0)  1=(2,1,0)
 \begin{equation}
 m_{\ell}= \left(  \begin{matrix}  \la^7 &    \la^6 &   \la^5 \\  \la^5 &    \la^4 &   \la^3 \\ \la^2 &   \la &   1 \end{matrix}     \right) \,,\qquad\qquad
 m_{\nu}= \left(  \begin{matrix}  \la^4 &    \la^3 &   \la^2 \\  \la^3 &   \la^2 &  \la \\ \la^2 &   \la &   1 \end{matrix}     \right)\,.
 \end{equation}
 \end{itemize}
 The values of the neutrino observables were computed extracting the modulus (argument) of the complex random coefficients 
 in the interval $[0.5,2]$ ($[0,2\pi]$) with a flat distribution. In order to ensure a reasonable hierarchy 
 for charged fermions, the values of $\lambda$ are different in the two cases: $\lambda=0.2$ for A,  $\lambda=0.4$ for H.
 The results of such a scan can be summarized as follows.
 Since the problem with Anarchy is that all mixing angles should be large and of the same order of magnitude, 
 it is quite difficult to reproduce $\theta_{13}$ of the order of the Cabibbo angle. In addition, the smallness of solar-to-atmospheric 
 mass hierarchy $r$ 
 is not easily reproduced, being generically one order of magnitude larger than expected.
 On the other hand, in the $H$ model  one can reproduce the correct size for $r$ and $\sin^2 \theta_{13}$, thus 
 making this option preferable over Anarchy.
 These results have been confirmed by a more recent analysis in \cite{melmer}.

\section{Summary and Outlook}

Neutrino physics deals with fundamental issues still of great importance, as the origin of masses and mixing and the possible 
existence of sterile neutrino species. In the domain of model building, an intense work to interpret the new data (especially after 
the recent measurement of $\theta_{13}$) has span a wide range of possibilities: beyond the ones illustrated here, one should also mention
the use of larger symmetries that already at LO  lead to non vanishing $\theta_{13}$ and non maximal $\theta_{23}$ \cite{altmix} to models where 
the flavor group and a generalized CP transformation are combined in a non trivial way \cite{cpfla,othercp}.
In spite of this huge effort, not a single model has been built which contains all features of the neutrino experimental data without
invoking some fine-tuning in the model parameters or ad-hoc constructions.
Thus, more work must be devoted to this field to clarify the origin of the present experimental data.

\vspace{.5cm}
\appendix{\bf Acknowledgments} 

I am very grateful to  the Organizers of the  QFTHEP'2015 workshop  for inviting me to give this review talk and for 
the beautiful and stimulating atmosphere during my stay in Samara.

\bibliographystyle{aipproc}

\begin{thebibliography}{99}

  
\bibitem{Abe:2011sj}
{\bf T2K} Collaboration, K.~Abe {\em et.~al.},  Phys. Rev. Lett. {\bf 107} (2011) 041801, arXiv:1106.2822;
  arXiv:1106.2822.
  
\bibitem{Adamson:2011qu}
{\bf MINOS} Collaboration, P.~Adamson {\em et.~al.},  Phys. Rev. Lett.
  {\bf 107} (2011) 181802, arXiv:1108.0015.

\bibitem{Abe:2011fz} 
  Y.~Abe {\it et al.} [Double Chooz Collaboration],
  Phys.\ Rev.\ Lett.\  {\bf 108}, 131801 (2012)
  doi:10.1103/PhysRevLett.108.131801
  [arXiv:1112.6353 [hep-ex]].

\bibitem{Ahn:2012nd} 
  J.~K.~Ahn {\it et al.} [RENO Collaboration],
  Phys.\ Rev.\ Lett.\  {\bf 108}, 191802 (2012)
  doi:10.1103/PhysRevLett.108.191802
  [arXiv:1204.0626 [hep-ex]].
  
  \bibitem{An:2012eh} 
  F.~P.~An {\it et al.} [Daya Bay Collaboration],
  Phys.\ Rev.\ Lett.\  {\bf 108}, 171803 (2012)
  doi:10.1103/PhysRevLett.108.171803
  [arXiv:1203.1669 [hep-ex]].

  \bibitem{Fogli:2012ua}
F.~Capozzi, G.~L.~Fogli, E.~Lisi, A.~Marrone, D.~Montanino and A.~Palazzo,
  Phys.\ Rev.\ D {\bf 89}, 093018 (2014)
  doi:10.1103/PhysRevD.89.093018
  [arXiv:1312.2878 [hep-ph]].

  \bibitem{GonzalezGarcia:2012}
M.~C.~Gonzalez-Garcia, M.~Maltoni and T.~Schwetz,
  arXiv:1512.06856 [hep-ph].

  \bibitem{Tortola:2012te}
  D.~V.~Forero, M.~Tortola and J.~W.~F.~Valle,
  Phys.\ Rev.\ D {\bf 90}, no. 9, 093006 (2014)
  doi:10.1103/PhysRevD.90.093006
  [arXiv:1405.7540 [hep-ph]].

\bibitem{Maki:1962mu} 
  B.~Pontecorvo,
  Sov.\ Phys.\ JETP {\bf 6}, 429 (1957)
  [Zh.\ Eksp.\ Teor.\ Fiz.\  {\bf 33}, 549 (1957)];\\
  Z.~Maki, M.~Nakagawa and S.~Sakata,
  Prog.\ Theor.\ Phys.\  {\bf 28}, 870 (1962).
  doi:10.1143/PTP.28.870.
  
  
\bibitem{Wolfenstein:1977ue} 
  L.~Wolfenstein,
  Phys.\ Rev.\ D {\bf 17}, 2369 (1978).
  
\bibitem{GonzalezGarcia:2002dz} 
  M.~C.~Gonzalez-Garcia and Y.~Nir,
  Rev.\ Mod.\ Phys.\  {\bf 75}, 345 (2003)
  [hep-ph/0202058].
  
\bibitem{prem}  
A. Dziewonski, Earth Structure, Global, in "The Encyclopedia of Solid Earth
Geophysics", ed. by D.E. James, (Van Nostrand Reinhold, New York, 1989);\\
R. Gandhi, C. Quigg, M. Hall Reno, and I. Sarcevic, Astroparticle Physics {\bf 5}, 81 (1996).  
  

  
  
  
\bibitem{Cervera:2000kp} 
  A.~Cervera, A.~Donini, M.~B.~Gavela, J.~J.~Gomez Cadenas, P.~Hernandez, O.~Mena and S.~Rigolin,
  Nucl.\ Phys.\ B {\bf 579}, 17 (2000)
  [Nucl.\ Phys.\ B {\bf 593}, 731 (2001)]
  [hep-ph/0002108].
  
\bibitem{Donini:2002rm} 
  A.~Donini, D.~Meloni and P.~Migliozzi,
  Nucl.\ Phys.\ B {\bf 646}, 321 (2002)
  [hep-ph/0206034].
  

\bibitem{white} K. Abazajian {\em et.~al.},  arXiv:1204.5379.  


\bibitem{Athanassopoulos:1995iw} 
  C.~Athanassopoulos {\it et al.} [LSND Collaboration],
  Phys.\ Rev.\ Lett.\  {\bf 75}, 2650 (1995)
  [nucl-ex/9504002].
  
 \bibitem{Donini:2001xy} 
  A.~Donini and D.~Meloni,
  Eur.\ Phys.\ J.\ C {\bf 22}, 179 (2001)
  doi:10.1007/s100520100777
  [hep-ph/0105089].
  
\bibitem{Maltoni:2002xd} 
  M.~Maltoni, T.~Schwetz, M.~A.~Tortola and J.~W.~F.~Valle,
  Nucl.\ Phys.\ B {\bf 643}, 321 (2002)
  doi:10.1016/S0550-3213(02)00747-2
  [hep-ph/0207157].
  

  
\bibitem{Mboo:2012}
 A.~A.~Aguilar-Arevalo {\it et al.} [MiniBooNE Collaboration],
  Phys.\ Rev.\ Lett.\  {\bf 110}, 161801 (2013)
  doi:10.1103/PhysRevLett.110.161801
  [arXiv:1207.4809 [hep-ex], arXiv:1303.2588 [hep-ex]]; Phys.\ Rev.\ Lett.\  {\bf 110}, 161801 (2013)
  doi:10.1103/PhysRevLett.110.161801
  [arXiv:1207.4809 [hep-ex], arXiv:1303.2588 [hep-ex]].

\bibitem{LSND} C.~Athanassopoulos {\it et al.} [LSND Collaboration],
  Phys.\ Rev.\ Lett.\  {\bf 75}, 2650 (1995)
  doi:10.1103/PhysRevLett.75.2650
  [nucl-ex/9504002]; Phys.\ Rev.\ Lett.\  {\bf 77}, 3082 (1996)
  doi:10.1103/PhysRevLett.77.3082
  [nucl-ex/9605003]; Phys.\ Rev.\ Lett.\  {\bf 81}, 1774 (1998)
  doi:10.1103/PhysRevLett.81.1774
  [nucl-ex/9709006].
  
  
\bibitem{Abdurashitov:1998ne} 
  J.~N.~Abdurashitov {\it et al.} [SAGE Collaboration],
  Phys.\ Rev.\ C {\bf 59}, 2246 (1999)
  [hep-ph/9803418].
 
 
\bibitem{Hampel:1997fc} 
  W.~Hampel {\it et al.} [GALLEX Collaboration],
  Phys.\ Lett.\ B {\bf 420}, 114 (1998).
 
 \bibitem{Acero:2007su} 
  M.~A.~Acero, C.~Giunti and M.~Laveder,
  Phys.\ Rev.\ D {\bf 78}, 073009 (2008)
  [arXiv:0711.4222 [hep-ph]].
  
\bibitem{Giunti:2010zu} 
  C.~Giunti and M.~Laveder,
  Phys.\ Rev.\ C {\bf 83}, 065504 (2011)
  [arXiv:1006.3244 [hep-ph]].
  
\bibitem{Mention:2011} 
  G.~Mention, M.~Fechner, T.~Lasserre, T.~A.~Mueller, D.~Lhuillier, M.~Cribier and A.~Letourneau,
  Phys.\ Rev.\ D {\bf 83}, 073006 (2011)
  doi:10.1103/PhysRevD.83.073006
  [arXiv:1101.2755 [hep-ex]].
  
\bibitem{Huber:2011} 
P.~Huber,
  Phys.\ Rev.\ C {\bf 84}, 024617 (2011)
  [Phys.\ Rev.\ C {\bf 85}, 029901 (2012)]
  doi:10.1103/PhysRevC.85.029901, 10.1103/PhysRevC.84.024617
  [arXiv:1106.0687 [hep-ph]].
   
\bibitem{Vogel:1980bk} 
  P.~Vogel, G.~K.~Schenter, F.~M.~Mann and R.~E.~Schenter,
  Phys.\ Rev.\ C {\bf 24}, 1543 (1981).
  
\bibitem{VonFeilitzsch:1982jw} 
  F.~Von Feilitzsch, A.~A.~Hahn and K.~Schreckenbach,
  Phys.\ Lett.\ B {\bf 118}, 162 (1982).  
  
  
\bibitem{Schreckenbach:1985ep} 
  K.~Schreckenbach, G.~Colvin, W.~Gelletly and F.~Von Feilitzsch,
  Phys.\ Lett.\ B {\bf 160}, 325 (1985).
  
\bibitem{Hahn:1989zr} 
  A.~A.~Hahn, K.~Schreckenbach, G.~Colvin, B.~Krusche, W.~Gelletly and F.~Von Feilitzsch,
  Phys.\ Lett.\ B {\bf 218}, 365 (1989).    
  
  
\bibitem{Declais:1994ma} 
  Y.~Declais {\it et al.},
  Phys.\ Lett.\ B {\bf 338}, 383 (1994).
  
 \bibitem{Kuvshinnikov:1990ry}
  A.~A.~Kuvshinnikov, L.~A.~Mikaelyan, S.~V.~Nikolaev, M.~D.~Skorokhvatov and A.~V.~Etenko,
  JETP Lett.\  {\bf 54} (1991) 253
   [Yad.\ Fiz.\  {\bf 52} (1990) 472]
   [Pisma Zh.\ Eksp.\ Teor.\ Fiz.\  {\bf 54} (1991) 259]
   [Sov.\ J.\ Nucl.\ Phys.\  {\bf 52} (1990) 300].
   
   
\bibitem{Boehm:2001ik}
  F.~Boehm {\it et al.},
  Phys.\ Rev.\ D {\bf 64} (2001) 112001
  [hep-ex/0107009].
  
\bibitem{Abe:2012tg} 
  Y.~Abe {\it et al.} [Double Chooz Collaboration],
  Phys.\ Rev.\ D {\bf 86}, 052008 (2012)
  [arXiv:1207.6632 [hep-ex]].  
  
\bibitem{Dydak:1983zq} 
  F.~Dydak {\it et al.},
  Phys.\ Lett.\ B {\bf 134}, 281 (1984).
  
\bibitem{Bilenky:1999ny}
  S.~M.~Bilenky, C.~Giunti, W.~Grimus and T.~Schwetz,
  Phys.\ Rev.\ D {\bf 60} (1999) 073007
  [hep-ph/9903454].  
  
\bibitem{Adamson:2010wi}
  P.~Adamson {\it et al.} [MINOS Collaboration],
  Phys.\ Rev.\ D {\bf 81} (2010) 052004
  [arXiv:1001.0336 [hep-ex]].  

\bibitem{Armbruster:2002mp} 
  B.~Armbruster {\it et al.} [KARMEN Collaboration],
  Phys.\ Rev.\ D {\bf 65}, 112001 (2002)
  [hep-ex/0203021].
   
\bibitem{Astier:2003gs} 
  P.~Astier {\it et al.} [NOMAD Collaboration],
  Phys.\ Lett.\ B {\bf 570}, 19 (2003)
  [hep-ex/0306037].      
   
 \bibitem{Antonello:2012}
 M.~Antonello {\it et al.},
  Eur.\ Phys.\ J.\ C {\bf 73}, no. 3, 2345 (2013)
  doi:10.1140/epjc/s10052-013-2345-6
  [arXiv:1209.0122 [hep-ex]].

\bibitem{agafo}
N.~Agafonova {\it et al.} [OPERA Collaboration],
  JHEP {\bf 1307}, 004 (2013)
  [JHEP {\bf 1307}, 085 (2013)]
  doi:10.1007/JHEP07(2013)004, 10.1007/JHEP07(2013)085
  [arXiv:1303.3953 [hep-ex]].
   
\bibitem{kopp} 
J.~Kopp, P.~A.~N.~Machado, M.~Maltoni and T.~Schwetz,
  JHEP {\bf 1305}, 050 (2013)
  doi:10.1007/JHEP05(2013)050
  [arXiv:1303.3011 [hep-ph]].
  
\bibitem{laveder} C.~Giunti, M.~Laveder, Y.~F.~Li and H.~W.~Long,
  Phys.\ Rev.\ D {\bf 88}, 073008 (2013)
  doi:10.1103/PhysRevD.88.073008
  [arXiv:1308.5288 [hep-ph]].
    
  
\bibitem{Harrison} P.~F. Harrison, D.~H. Perkins, and W.~G. Scott,  Phys. Lett. {\bf B530} (2002) 167, 
arXiv:hep-ph/0202074; P.~F. Harrison and W.~G. Scott, ,  Phys. Lett. {\bf B535} (2002) 163,
arXiv:hep-ph/0203209;
Z.-z. Xing, Phys. Lett. {\bf B533} (2002) 85, arXiv:hep-ph/0204049;
P.~F. Harrison and W.~G. Scott,  Phys. Lett. {\bf B547} (2002) 219,
arXiv:hep-ph/0210197;  Phys. Lett. {\bf B557}(2003) 76, arXiv:hep-ph/0302025.

\bibitem{GR1}
Y.~Kajiyama, M.~Raidal, and A.~Strumia,  Phys. Rev. {\bf D76} (2007) 117301,arXiv:0705.4559; L.~L. Everett and A.~J. Stuart,  Phys. Rev. {\bf
  D79} (2009) 085005, arXiv:0812.1057; G.-J. Ding, L.~L. Everett, and A.~J. Stuart,  Nucl. Phys. {\bf B857} (2012) 219, arXiv:1110.1688;
F.~Feruglio and A.~Paris,   JHEP {\bf 03} (2011) 101, arXiv:1101.0393.

\bibitem{Altarelli:2010gt} 
  G.~Altarelli and F.~Feruglio,
  Rev.\ Mod.\ Phys.\  {\bf 82}, 2701 (2010)
  [arXiv:1002.0211 [hep-ph]].
  
\bibitem{Altarelli:2006ty}
G.~Altarelli and F.~Feruglio,  Nucl. Phys.  {\bf B741} (2006 ) 215, arXiv:hep-ph/0512103;
G.~Altarelli and D.~Meloni,
J.\ Phys.\ G {\bf 36}, 085005 (2009)
[arXiv:0905.0620 [hep-ph]]. 


\bibitem{Altarelli:2009gn} 
  G.~Altarelli, F.~Feruglio and L.~Merlo,
  JHEP {\bf 0905}, 020 (2009)
  [arXiv:0903.1940 [hep-ph]].
  
  
  
\bibitem{Meloni:2011fx}
  D.~Meloni,
  JHEP {\bf 1110} (2011) 010
  [arXiv:1107.0221 [hep-ph]].


 
\bibitem{Raidal:2004iw} 
  M.~Raidal,
  Phys.\ Rev.\ Lett.\  {\bf 93}, 161801 (2004)
  [hep-ph/0404046].
  
  
 \bibitem{Antusch:2005ca} 
  S.~Antusch, S.~F.~King and R.~N.~Mohapatra,
  Phys.\ Lett.\ B {\bf 618}, 150 (2005)
  [hep-ph/0504007]. 
  
 
 
\bibitem{Altarelli:2015foa}
  G.~Altarelli, P.~A.~N.~Machado and D.~Meloni,
  PoS CORFU {\bf 2014} (2015) 012
  [arXiv:1504.05514 [hep-ph]].     
 

\bibitem{blankenburg}
G.~Altarelli and G.~Blankenburg,
  JHEP {\bf 1103} (2011) 133
  [arXiv:1012.2697 [hep-ph]].

\bibitem{Froggatt:1978nt}
  C.~D.~Froggatt and H.~B.~Nielsen,
  Nucl.\ Phys.\ B {\bf 147} (1979) 277.



\bibitem{5DSU5}
E.~Witten,
  Nucl.\ Phys.\  B {\bf 258} (1985) 75;
Y.~Kawamura,
  Prog.\ Theor.\ Phys.\  {\bf 105} (2001) 999
  [arXiv:hep-ph/0012125];
A.~E.~Faraggi,
  Phys.\ Lett.\  B {\bf 520} (2001) 337
  [arXiv:hep-ph/0107094] and references therein.

\bibitem{5D}
L.~J.~Hall and Y.~Nomura,
  Phys.\ Rev.\  D {\bf 64} (2001) 055003
  [arXiv:hep-ph/0103125];
Y.~Nomura,
  Phys.\ Rev.\  D {\bf 65} (2002) 085036
  [arXiv:hep-ph/0108170];
L.~J.~Hall and Y.~Nomura,
  Phys.\ Rev.\  D {\bf 66} (2002) 075004
  [arXiv:hep-ph/0205067].
  
\bibitem{5Dfreedom}
G.~Altarelli and F.~Feruglio,
  Phys.\ Lett.\  B {\bf 511} (2001) 257
  [arXiv:hep-ph/0102301];
A.~Hebecker and J.~March-Russell,
  Nucl.\ Phys.\  B {\bf 613} (2001) 3
  [arXiv:hep-ph/0106166];
A.~Hebecker and J.~March-Russell,
  Phys.\ Lett.\  B {\bf 541} (2002) 338
  [arXiv:hep-ph/0205143].

\bibitem{Altarelli:2008bg}
  G.~Altarelli, F.~Feruglio, C.~Hagedorn,
  JHEP {\bf 0803}, 052-052 (2008).
  [arXiv:0802.0090 [hep-ph]].     
  
\bibitem{Hall:1999sn}
 L.~J. Hall, H.~Murayama, and N.~Weiner,  Phys.
   Rev. Lett. {\bf 84} (2000) 2572,  arXiv:hep-ph/9911341.

\bibitem{deGouvea:2003xe}
A.~de~Gouvea and H.~Murayama,  Phys. Lett.
  {\bf B573} (2003) 94, arXiv:hep-ph/0301050.  
  
  
 \bibitem{Altarelli:2012ia} 
  G.~Altarelli, F.~Feruglio, I.~Masina and L.~Merlo,
  JHEP {\bf 1211}, 139 (2012)
  [arXiv:1207.0587 [hep-ph]]. 
  


\bibitem{deGouvea:2012ac} 
  A.~de Gouvea and H.~Murayama,
  Phys.\ Lett.\ B {\bf 747}, 479 (2015)
  doi:10.1016/j.physletb.2015.06.028
  [arXiv:1204.1249 [hep-ph]].


\bibitem{melmer} J.~Bergstrom, D.~Meloni and L.~Merlo,
  Phys.\ Rev.\ D {\bf 89} (2014) 9,  093021
  [arXiv:1403.4528 [hep-ph]].


\bibitem{altmix} R.~d.~A.~Toorop, F.~Feruglio and C.~Hagedorn,
Phys.\ Lett.\ B {\bf 703} (2011) 447 [arXiv:1107.3486 [hep-ph]]; Nucl.\ Phys.\ B{\bf 858}, 437 (2012) , arXiv:1112.1340; 
S.~F.~King, C.~Luhn and A.~J.~Stuart,
  Nucl.\ Phys.\ B {\bf 867} (2013) 203
  [arXiv:1207.5741 [hep-ph]]; 
C.~Hagedorn and D.~Meloni,
  Nucl.\ Phys.\ B {\bf 862} (2012) 691
  [arXiv:1204.0715 [hep-ph]];   
S.~F.~King, T.~Neder and A.~J.~Stuart,
  Phys.\ Lett.\ B {\bf 726} (2013) 312
  [arXiv:1305.3200 [hep-ph]];  



\bibitem{cpfla} 
G.~Ecker, W.~Grimus and H.~Neufeld,
J.\ Phys.\ A {\bf 20} (1987) L807;  Int.\ J.\ Mod.\ Phys.\ A {\bf 3} (1988) 603;
W.~Grimus and M.~N.~Rebelo,
Phys.\ Rept.\  {\bf 281} (1997) 239
[hep-ph/9506272];
W.~Grimus and L.~Lavoura,
Phys.\ Lett.\ B {\bf 579} (2004) 113 [hep-ph/0305309];
M.~Holthausen, M.~Lindner and M.~A.~Schmidt,
JHEP {\bf 1304} (2013) 122 [arXiv:1211.6953 [hep-ph]];
F.~Feruglio, C.~Hagedorn and R.~Ziegler,
JHEP {\bf 1307} (2013) 027
[arXiv:1211.5560 [hep-ph]]; Eur.\ Phys.\ J.\ C {\bf 74} (2014) 2753
[arXiv:1303.7178 [hep-ph]].

\bibitem{othercp} 
I.~de Medeiros Varzielas and D.~Emmanuel-Costa,
  Phys.\ Rev.\ D {\bf 84} (2011) 117901
  [arXiv:1106.5477 [hep-ph]];
G.~Bhattacharyya, I.~de Medeiros Varzielas and P.~Leser,
Phys.\ Rev.\ Lett.\  {\bf 109} (2012) 241603
[arXiv:1210.0545 [hep-ph]]; 
M.~C.~Chen and K.~T.~Mahanthappa,
Phys.\ Lett.\ B {\bf 681} (2009) 444
[arXiv:0904.1721 [hep-ph]]; 
 L.~L.~Everett, T.~Garon and A.~J.~Stuart,
  JHEP {\bf 1504}, 069 (2015)
  doi:10.1007/JHEP04(2015)069
  [arXiv:1501.04336 [hep-ph]];
C.~C.~Li and G.~J.~Ding,
  JHEP {\bf 1505}, 100 (2015)
  doi:10.1007/JHEP05(2015)100
  [arXiv:1503.03711 [hep-ph]];
A.~Di Iura, C.~Hagedorn and D.~Meloni,
  JHEP {\bf 1508}, 037 (2015)
  doi:10.1007/JHEP08(2015)037
  [arXiv:1503.04140 [hep-ph]].

 
\end{thebibliography}

\end{document}